\documentclass[aps,pra,twocolumn,groupedaddress,superscriptaddress,floatfix,showpacs]{revtex4-1}
\usepackage{graphicx}% Include figure files
\usepackage{dcolumn}% Align table columns on decimal point
\usepackage{bm}% bold math
\usepackage{natbib}
\usepackage{amsmath}
\begin{document}
\title{Quantum interference and light polarization effects in unresolvable atomic lines:
~application to a precise measurement of the $^{6,7}$Li D2 lines}
\date{\today}
\author{Roger C. Brown}  %\altaffiliation[Electronic address: ]{ robrown@umd.edu}
\affiliation{National Institute of Standards and Technology,
Gaithersburg, MD  20899}
\affiliation{University of Maryland, College Park, MD  20742}
\author{Saijun Wu}
\altaffiliation{Current address: Department of Physics, College of
Science, Swansea University, Swansea, SA2 8PP, United Kingdom}
\affiliation{National Institute of Standards and Technology,
Gaithersburg, MD  20899}
\affiliation{University of Maryland, College Park, MD  20742}
\author{J. V. Porto} %\affiliation{University of Maryland, College Park, MD  20742}
\affiliation{National Institute of Standards and Technology,
Gaithersburg, MD  20899}
\affiliation{University of Maryland, College Park, MD  20742}
\author{Craig J. Sansonetti}
\affiliation{National Institute of Standards and Technology,
Gaithersburg, MD  20899}
\author{C.E. Simien}
\altaffiliation{Current address: Department of Physics,  West Virginia University, Morgantown, WV 26506}
\author{Samuel M. Brewer}
\altaffiliation{Current address: Ion Storage Group,Time and Frequency Division NIST, 325 Broadway Boulder, CO 80305}
\affiliation{National Institute of Standards and Technology,
Gaithersburg, MD  20899}
\affiliation{University of Maryland, College Park, MD  20742}
\author{Joseph N. Tan}
\author{J. D. Gillaspy}
\affiliation{National Institute of Standards and Technology,
Gaithersburg, MD  20899}
\begin{abstract}
We characterize the effect of quantum interference on the line shapes and measured line positions in atomic spectra. These effects, which occur when the excited state splittings are of order the natural line widths, represent an overlooked but significant systematic effect. We show that excited state interference gives rise to non-Lorenztian line shapes that depend on excitation polarization, and we present expressions for the corrected line shapes. We present spectra of  $^{6,7}$Li D lines taken at multiple excitation laser polarizations and show that failure to account for interference changes the inferred line strengths and shifts the line centers by as much as 1~MHz. Using the correct line shape, we determine absolute optical transition frequencies with an uncertainty of $\leq$25~kHz and provide an improved determination of the difference in mean square nuclear charge radii between $^{6}$Li and $^{7}$Li.  This analysis should be important for a number of high resolution spectral measurements that include partially resolvable atomic lines.
\end{abstract}

\maketitle

\section{Introduction}
The measurement of accurate atomic transition frequencies plays an important role in fundamental physics from atomic clocks to the determination of nuclear charge radii.  %\cite{Demtroder,LatticeClockRMP2011}.
Determining accurate frequencies requires a sufficient understanding of the transition line shape.  In particular, the Lorentzian line shape %(or Voigt profile when convolved with a Gaussian)
is of fundamental importance in the analysis of resonant phenomena in many areas of physics~\cite{wignerweiss}.  When two or more resonances are separated on the order of a natural line width, unresolvable in a fundamental sense not limited by instrumentation, there arises the possibility of interference. The resulting line shape is, in general, no longer a simple sum of Lorentzians, even in the low intensity limit.  Although this effect has been known in different contexts for many years~\cite{FrankenPhysRev1961,BergemanJChemPhys1974,Walkup1982,Hessels2010}, it has typically been ignored in the interpretation of Doppler free spectra. In our previous work~\cite{Sansonetti2011,*Sansonetti2011erratum}, we demonstrated that quantum interference has an observable effect on atomic spectra, which can limit accuracy if not properly accounted for.  In section~\ref{sec:theory} of this article, we derive a more general set of line-shapes and estimate the systematic errors incurred if strictly Lorentzian line shapes are assumed.  In section~\ref{sec:experiment}, we use the more complete line shapes to extract absolute optical transition frequencies from new experimental $^{6,7}$Li data and quantify errors associated with incomplete line shapes. Finally, in section~\ref{sec:ncr}, we use our new measurement of the $^{6,7}$Li D line isotope shift to extract the relative $^{6,7}$Li difference in mean square nuclear charge radius.
The unresolvable hyperfine structure in the D2 lines of hydrogen~\cite{Eikema2001}, lithium~\cite{Sansonetti2011, *Sansonetti2011erratum}, potassium~\cite{Falke}, francium~\cite{CocFr1985}, singly-ionized beryllium~\cite{BeNucChargeRadiJPG2010} and magnesium~\cite{MukherjeeEPJD2005} are additional examples where interference modified Lorentzian line shapes are expected.
%In a series of experiments, it was observed that the polarization of scattered light changed from polarized with the excitation polarization far from atomic resonance to strongly depolarized near resonance ~\cite{Tam1976,Kroop1978,Hamaguchi1980,Meng1992}. This effect is qualitatively explained in terms of a change from $\vec{J}=\vec{L}+\vec{S}$ total electronic angular momentum selection rules far from resonance to $\vec{F}=\vec{I}+\vec{J}$ angular momentum coupling near resonance, where $\vec{L}$ is the electronic angular momentum, $\vec{S}$ is the electron spin angular momentum, and $\vec{I}$ is the nuclear angular momentum.  It was further noted that this effect, in the case of Doppler free spectra of closely spaced resonances, would manifest itself as additional quantum interference terms in the line shape.
\section{Dipole Scattering line shape}
\label{sec:theory}
%Probably should also cite:
%The Effect of Hyperfine Structure Due to Nuclear Spin on Polarization of Resonance Radiation ~\cite{Ellett1930}
%Light Scattering in Terms of Oscillator Strengths and Refractive Indices ~\cite{Penney1969}
%the context of frequency dependent polarization light scattering

We begin with a derivation of the corrected line shape, including quantum interference terms, using the Kramers-Heisenberg formula~\cite{Loudon} which describes the differential scattering rate of light incident on an atom initially in the state $|i\rangle$ and ending in the state $|f\rangle$. It can be derived from Fermi's golden rule~\cite{FermiGoldenRuleDirac1927}
\begin{equation}
\frac{d R_{i \rightarrow f}}{d \Omega_s} =
\frac{2 \pi}{\hbar} \left| M_{f i}\right|^2 \rho_s, \label{eq:Fgolden}
\end{equation}
where $\hbar$ is Plank's constant($h$) divided by $2\pi$ and $\rho_s$ is the density of scattered photon states into a solid angle $d\Omega_s$ along the scattering direction ${\bf k}_s$. The scattering matrix element $M_{fi}$ is calculated to second order in the electric dipole coupling.  The scattering matrix element depends on the frequency, wavevector and polarization of the incident light ($\omega_{\rm L}, {\bf k}_{\rm L}, \hat{\epsilon}_{\rm L}$) and scattered light ($\omega_s, {\bf k}_s, \hat{\epsilon}_s$). The resulting scattering rate is:
\begin{equation}
\frac{d R_{i \rightarrow f}}{d \Omega_s} =
\frac{  \pi E_{\rm L}^2 \omega_s^3}{ h^3 c^3 \epsilon_0} \left|
	 \sum_j \frac{ (\hat{\epsilon}_s^* \cdot  {\bf D}_{fj} )\  ( {\bf D}_{j i}   \cdot \hat{\epsilon}_{\rm L})}
	 { \omega_{j i} - \omega_{\rm L} - i \Gamma_j/2} \right|^2
	 \label{eq:HK}
\end{equation}
where c is the speed of light, $\epsilon_0$ is the permittivity of free space, and $E_{\rm L}$ is the amplitude of the electric field of the incident light.
The sum is over excited intermediate states $|j\rangle$ with transition frequencies $\omega_{ji}$ and atomic dipole matrix elements ${\bf D}_{j i} = \langle j | e {\bf r} | i\rangle$. Here $e$ is the electron charge and ${\bf r}$ is the position operator of the valence electron. The finite lifetime of the excited states $|j\rangle$ are accounted for~\cite{Loudon} by including the imaginary part $i \Gamma_j/2$ in the transition frequency $\omega_{ji}$ \footnote{One may consider whether the interference effects described in this paper could also modify the simple replacement $ \omega_{ji} \rightarrow  \omega_{ji} + i \Gamma_j /2$  when accounting for the coupling to the continuum. However, in the cases considered here (a single electronic state split by fine and hyperfine structure), the effect of interference disappears when integrated over all solid angle. Since the inclusion of $\Gamma_j$ results from summing the coupling to the continuum over all solid angle, it is probable that the addition of $i \Gamma_j /2$ correctly accounts for the continuum, although a more detailed calculation would be needed to confirm this. Empirically, the line shapes presented here well fit the observed data}. Here $\Gamma_j$ is the inverse lifetime (or full width half maximum for an isolated Lorentzian line) of $|j\rangle$.
%and characterizes the coupling strength of the atomic excited state to the electromagnetic continuum.  What is different in this work is the presence of nearby excited states, not the form of the coupling or the nature of the continuum.  This is experimentally justified by the observed Lorenzian line-shapes of the resolvable D1 lines.
Equation~\ref{eq:HK}, valid in the low excitation intensity limit, does not include multiple scattering effects like optical pumping. Additionally, we make the rotating wave approximation, which is appropriate for near resonant excitation. While Eq.~\ref{eq:HK} is a Lorentzian distribution if only one term of the sum is considered, since the sum over intermediate states is {\em inside} the square, one can see that interference from different excited states $|j\rangle$ is possible. %, which can affect nearby transitions.

For a concrete experimental comparison, we restrict our analysis to the case where states $| i \rangle$ and $| f \rangle$ are hyperfine states of a single electronic ground state with electronic angular momentum $J$, and the intermediate hyperfine states $| j \rangle$ belong to a single excited electronic state with angular momentum $J^\prime$.  The states are labeled by their total angular momentum and z-projection of angular momentum $|F_i, m_i\rangle$, $|F_f, m_f\rangle$, and $|F^\prime, m^\prime \rangle$.

One can evaluate the atom field coupling matrix element by repeatedly applying the Wigner-Eckhart theorem. The reduced matrix elements can be written in terms of the electronic excited state linewidth $\Gamma$ and a reference intensity $I_0$ (see Appendix~\ref{sec:CGappendix}). (For a closed transition such as the Li 2s-2p transitions considered here,  $\Gamma = \Gamma_j$.) This gives
\begin{equation}
\frac{d R_{i \rightarrow f}}{d \Omega_s} = \frac{3}{8 \pi}
  \frac{I}{I_0}
	\left(\frac{\Gamma}{2}  \right)^{3}
 	\left|
	\sum_{F^\prime m^\prime}
	  \frac{ (\hat{\epsilon}_s \cdot  {\bf A}^{ F^\prime m^\prime}_{F_f m_f} )\
	  ( {\bf A}^{ F^\prime m^\prime}_{F_i m_i}   \cdot \hat{\epsilon}_{\rm L})}
	 {\Delta^{ F^\prime}_{F_i}  + i \Gamma/2}  \right|^2. \label{eq:lineshape1}
\end{equation}
Here $\Delta^{F^\prime}_{F_i} = \omega_{\rm L}-\omega_{F^\prime F_i}$, and ${\bf A}^{ F^\prime m^\prime}_{F_f m_f}$ are the normalized dipole matrix elements containing all the angular dependence of the atomic dipole. The explicit form for ${\bf A}^{ F^\prime m^\prime}_{F_i m_i}$ is given in Appendix~\ref{sec:CGappendix}.

Since the denominator in Eq.~\ref{eq:lineshape1} is independent of $m^\prime$, we can sum the numerator over $m^\prime$. Defining the function
\begin{equation}
C^{F^\prime}_{i\rightarrow f}(\hat{\epsilon}_s, \hat{\epsilon}_{\rm L})=\sum_{m^\prime}(\hat{\epsilon}_s \cdot  {\bf A}^{ F^\prime m^\prime}_{F_f m_f} )  ( {\bf A}^{ F^\prime m^\prime}_{F_i m_i}   \cdot \hat{\epsilon}_{\rm L}),
\label{eq:4}
\end{equation}
we have
\begin{equation}
\frac{d R_{i \rightarrow f}}{d \Omega_s} = \frac{3}{8 \pi}
  \frac{I}{I_0}
	\left(\frac{\Gamma}{2}  \right)^{3}
 	\left|
	\sum_{F^\prime}
	  \frac{ C^{F^\prime}_{i\rightarrow f}(\hat{\epsilon}_s, \hat{\epsilon}_{\rm L})}
	 {\Delta^{ F^\prime}_{F_i}  + i \Gamma/2}  \right|^2, \label{eq:lineshape2}
\end{equation}
where $C^{F^\prime}_{i\rightarrow f}(\hat{\epsilon}_s, \hat{\epsilon}_{\rm L})$ depends on the initial and final state quantum numbers $F_i$, $m_i$, $F_f$ and $m_f$.

Equation~\ref{eq:lineshape2} describes the differential scattering rate of light into solid angle $d \Omega_s$~(along ${\bf k}_s$) with polarization $\hat{\epsilon}_s$ for atoms starting in state $|F_i, m_i\rangle$ and ending in $|F_f,m_f\rangle$.  In a typical spectroscopy experiment, the final scattering state is unresolved, so the scattering rate $R_{F_i m_i \rightarrow F_f m_f}$ is summed over final states $F_f$ and $m_f$. To further simplify the discussion, we assume the detection is polarization insensitive and sum over the two scattered polarizations $\hat{\epsilon}_s \perp {\bf k}_s$ for a given detection direction ${\bf k}_s$.  If, in addition, we assume an unpolarized atomic sample, we must average over all initial $m_i$. Summing and evaluating the square in Eq.~\ref{eq:lineshape2}, gives rise to sums of Lorentzian components and cross-terms
%\frac{3}{8 \pi} check this
\begin{eqnarray}
\frac{d R_{F_i}(\hat{\epsilon}_{\rm L})}{d \Omega_s} & = &
\frac{1}{4 \pi}\frac{I}{I_0}
	\left(\frac{\Gamma}{2}  \right)^{3}
	\left(
	\sum_{F^\prime}
	  \frac{  f({\bf k}_s, \hat{\epsilon}_{\rm L}, F_i, F^\prime) }
	 {( \Delta^{F^\prime}_{F_i})^2 + ( \Gamma/2)^2} + \right.  \\
	 & &  \left.
		\sum_{F^\prime\neq F^{\prime \prime}} 2 Re \left[
	  \frac{ g({\bf k}_s, \hat{\epsilon}_{\rm L}, F_i, F^\prime,F^{\prime \prime})  }
	 {( \Delta^{ F^\prime}_{F_i} + i \Gamma/2)( \Delta^{ F^{\prime \prime}}_{F_i} - i \Gamma/2) } \right] \right), \nonumber
	 \label{eq:lineshape3}
\end{eqnarray}
where the line strengths $ f({\bf k}_s, \hat{\epsilon}_{\rm L}, F_i, F^\prime)$ and cross-term strengths $g({\bf k}_s, \hat{\epsilon}_{\rm L}, F_i, F^\prime,F^{\prime \prime})$ for a particular laser polarization and detected direction are given by
\begin{widetext}
\begin{eqnarray}
f({\bf k}_s, \hat{\epsilon}_{\rm L}, F_i, F^\prime)& = &\frac{3}{2 g_{\mathrm T}} \sum_{s, m_i F_f m_f} \left| C^{F^\prime}_{i\rightarrow f}(\hat{\epsilon}_s, \hat{\epsilon}_{\rm L}) \right|^2  \nonumber \\
g({\bf k}_s, \hat{\epsilon}_{\rm L}, F_i, F^\prime,F^{\prime \prime})&=&\frac{3}{2 g_{\mathrm T}}\sum_{s, m_i F_f m_f} C^{F^\prime}_{i\rightarrow f}(\hat{\epsilon}_s, \hat{\epsilon}_{\rm L}) \left[ C^{F^{\prime \prime}}_{i\rightarrow f}(\hat{\epsilon}_s, \hat{\epsilon}_{\rm L}) \right]^* ,
\label{eq:fandg}
\end{eqnarray}
\end{widetext}
where $g_{\mathrm T} = \sum_i (2 F_i +1)$ is the total number of Zeeman states in the ground electronic state, assumed here to be uniformly thermally populated. When the excited state hyperfine splitting is not well resolved, $\Delta^{ F^\prime}_{F_i}-\Delta^{F^{\prime \prime}}_{F_i}\equiv \Delta^{F^{\prime \prime}}_{F^{\prime}} \approx \Gamma$, then the cross-terms are not necessarily negligible, as implicitly assumed in the latter portion of~\cite{Kielkopf73}.

\begin{figure}[h]
\centering\includegraphics [width=2 in,angle=0] {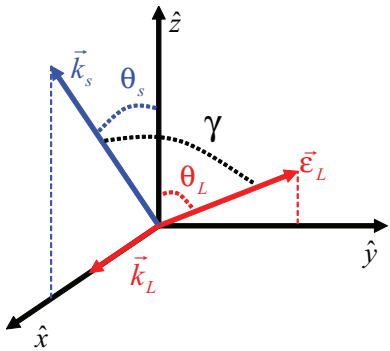}
\caption{(color online). Coordinate system: The excitation laser propagates along $\hat{x}$, so that the linear polarization direction $\hat{\epsilon}_{\rm L}$ lies in the $\hat{y}$-$\hat{z}$ plane, parameterized by $\theta_{\rm L}$.  The detection direction $k_s$ lies in the $\hat{x}$-$\hat{z}$ plane and is parameterized by $\theta_{\rm{s}}$. In our apparatus light collection is centered along $\hat{z}$ with an angular spread determined by the numerical aperture of the imaging system. The atomic beam is along $\hat{y}$.}%~(perpendicular to the excitation direction for Doppler-free spectroscopy).}
  %the z axis is the detection direction, the excitation laser propagates along the x direction, and the atomic beam propagates along the y direction. $\varphi_s$ and $\gamma$ represents the angle between the excitation polarization and the scattered light.
\label{fig:coordinatesystem}
\end{figure}

\subsection{Angular dependence}
\label{Angulardependence}
Dipole scattering of light follows a dipole radiation pattern~\cite{corney77}, which for linearly polarized light depends only on the angle $\gamma$ between excitation laser polarization~$\epsilon_{\rm L}$ and the fluorescence collection direction~${\bf k}_s$. The angular dependence of the dipole scattering is proportional to $\cos^2{\gamma}$, and it can always be written as a sum of a spherically symmetric component and a dipole component $(A_{\rm{tot}} + B P_2(\cos{\gamma }) )/4 \pi$. Here $A_{\rm{tot}}$ is the total line strength integrated over all solid angle, $P_2(x)=(3x^2-1)/2$ is the second Legendre polynomial (which has zero integral over solid angle), and $B$ characterizes the amplitude of the angular dependence. By construction $f$ contains all the scattering linestrength, the integral of the cross-terms $g$, proportional to $P_2(\cos{\gamma })$, over solid angle vanishes.  A consequence of this angular dependence is that $f({\bf k}_s, \hat{\epsilon}_{\rm L}, F_i, F^\prime)$ does {\em not} provide the correct ratio of line strengths of the $F_i \rightarrow F^\prime$ transitions for an arbitrary choice of detection direction, $\gamma$, since $B/A_{\rm{tot}}$ is not the same for different $F^\prime$. As we will show, however, there exist ``magic" orientations where $f$ does give line strengths consistent with resolved transitions. More importantly, at these magic conditions the cross-terms $g$ vanish, giving rise to purely Lorentzian line shapes.

%Since there is no external field to provide an independent quantization axis, the experimental geometry is completely defined by the excitation laser polarization and the fluorescence collection direction.  A further experimental constraint for the spectra to be Doppler free is that the atomic beam must be orthogonal to the excitation laser. However, for an unpolarized atomic beam this doesn't affect the angular distribution of the scattered light and may be regarded as less fundamental.

We parameterize $\gamma$ in terms of angles relevant to an experimental geometry.
The wave vectors $\hat{k}_{\rm L}$ and $\hat{k}_s$ define a plane which we take to be the $\hat{x}$-$\hat{z}$ plane.  Without loss of generality we can take $\hat{k}_{\rm L}$ along $\hat{x}$, so that $\hat{\epsilon}_{\rm L}$ lies in the $\hat{y}$-$\hat{z}$ plane, making an angle $\theta_{\rm L}$ with respect to $\hat{z}$, and $\hat{k}_s$ lies in the $\hat{x}$-$\hat{z}$ plane making an angle $\theta_{\rm{s}}$ with respect to $\hat{z}$, see Fig.~\ref{fig:coordinatesystem}.
The scattering is then parameterized by the linearly independent angles $\theta_{\rm{s}}$ and $\theta_{\rm L}$;   $f({\bf k}_s, \hat{\epsilon}_{\rm L}, F_i, F^\prime) = f(\theta_{\rm{s}}, \theta_{\rm L}, F_i, F^\prime)$ and similarly for $g$ \footnote{Alternatively, $\hat{\epsilon}_{\rm L}$ could be fixed along $\hat{z}$, and ${\bf k}_s$ could be characterized by polar angles~$(\theta^\prime_s, \phi^\prime_s)$. This would have the advantage that $\theta^\prime_s = \gamma$, but it is more convenient for a fixed scattering geometry to have $\theta_{\rm L}$ be one of the free parameters.}. The spherical harmonic addition theorem~\cite{Arfken} can be used to relate $P_2(\cos{\gamma})$ to $\theta_{\rm{s}}$ and $\theta_{\rm L}$:
\begin{equation}
P_2(\cos{\gamma}) = \frac{1}{2}\left(3 \cos^2{\theta_{\rm{s}}} \cos^2{\theta_{\rm L}} -1\right).
\end{equation}
The general form for $f$ and $g$ is then
\begin{eqnarray}
f(\theta_{\rm{s}},\theta_{\rm L}, F,F^\prime) & = & A^{F^\prime}_{F} +  \frac{B^{F^\prime}_{F} }{2}\left(3 \cos^2{\theta_{\rm{s}}} \cos^2{\theta_{\rm L}} -1\right) \nonumber\\
g(\theta_{\rm{s}},\theta_{\rm L}, F,F^\prime,F^{\prime \prime}) &=&   \frac{C^{F^\prime F^{\prime \prime}}_{F}}{2}\left(3 \cos^2{\theta_{\rm{s}}} \cos^2{\theta_{\rm L}} -1\right),
\label{eq:fandgoftheta}
\end{eqnarray}
where $A^{F^\prime}_{F}$, $B^{F^\prime}_{F} $ and $C^{F^\prime F^{\prime \prime}}_{F}$ are constants determined by evaluating Eq.\ref{eq:fandg}.
When $ \cos{\theta_{\rm{s}}} \cos{\theta_{\rm L}} =1/\sqrt{3}$, $g$ vanishes and $f$ correctly gives the line strength ratios. This can occur for a range of geometries. In particular, when the detection ${\bf k}_s$ is orthogonal to the excitation ${\bf k}_{\rm L}$~($\theta_{\rm L} = \gamma$,~$\theta_{\rm{s}} = 0$) as in our apparatus~\cite{Sansonetti2011, *Sansonetti2011erratum}, then $\theta_{\rm L}= \arccos(\frac{1}{\sqrt{3}}) \equiv \theta_{\rm{M}} \approx 54.73 ^{\rm{o}} $ is the so called ``magic" angle. Similar magic angle effects occur in quantum beat spectroscopy, which could be viewed as a time domain analogue of the effect considered here, where the excitation pulse width replaces the natural width~\cite{QuantumBeat1975,Demtroder}.  Explicit expressions for $f(\theta_{\rm L},F_i \rightarrow F^\prime)$ and $g(\theta_{\rm L},F_i \rightarrow F^\prime, F_i \rightarrow F'')$ are evaluated for lithium with the collection along the $\hat{z}$ direction in Appendix~\ref{sec:tablesappendix}.

\subsection{Line shape impact on extracted frequencies}
\label{lineshapeImpactonExtractedFrequencies}
\begin{figure}[h]
\centering\includegraphics [width=3.3 in,angle=0] {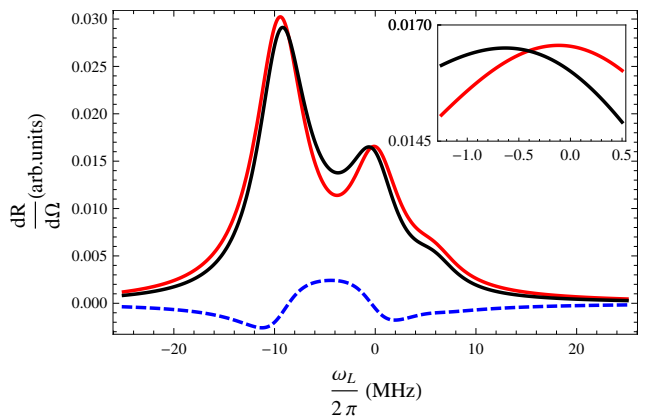}
\caption{(color online). The scattering rate (or intensity), $\frac{dR}{d\Omega}$ in arbitrary units, of the $F=2 \rightarrow F^\prime=1,2,3$ Doppler-free feature in $^7$Li with $\theta_{\rm L}=0$. Red: sum of Lorentzians with polarization independent weights, Blue dashed: sum of cross-terms, Black: sum of Lorentzians and cross-terms.  The laser frequency,~$\omega_{\rm L}/2\pi$, is the offset from the $F=2 \rightarrow F^\prime=2$ peak in units of MHz.  The inset is the F=2 $F=2 \rightarrow F^\prime=2$ peak enlarged to show the shift in line center.}
\label{lineshape}
\end{figure}

We now give a qualitative discussion of the effect of the additional interference terms on Doppler-free, or nearly free, spectra.  We choose $^{6,7}$Li as an example because of its fundamentally unresolvable structure~($\Delta^{F^{\prime \prime}}_{F^{\prime}}/\Gamma \approx 1$) and because it allows for direct comparison to experimental data.  Fig~\ref{lineshape} illustrates two primary effects.  First, the maxima of the total line shape are shifted relative to what is predicted by a simple sum of Lorentzian distributions, which can lead to errors in extracting the weighted line center.  Second, peaks may vary in intensity and prominence depending on the polarization angle of the laser.  For example in Fig~\ref{lineshape}, $\theta_{\rm L} = 0$, the amplitude of the $F=2 \rightarrow F^\prime=3$ component is reduced with respect to the $F=2 \rightarrow F^\prime=2$ component.
%which on first glance may lead one to think that their experimental data lacks suitable signal to noise to properly resolve the components.

Line centers are typically determined by fitting a sum of Lorentzian functions to the observed spectral profile.  We characterize the effect of cross-terms on line centers (both of individual hyperfine components and of centers of gravity of composite features) by taking a Doppler-free line shape given by Eq.~\ref{eq:lineshape3} with cross-terms and fitting to it using only Lorentzian functions~(amplitude, center, offset, linewidth). We then compare the centers given by Eq.~\ref{eq:lineshape3} to the centers extracted from the fit to estimate the effect of the cross-terms on measured quantities. From Eq.~\ref{eq:fandgoftheta}~(with $\theta_{\rm L} = \gamma$,~$\theta_{\rm{s}}=0$), one can see that the magnitude of the shifts, proportional to the angular dependent terms, has maxima at $\theta_{\rm L}=0,\pi/2$ and the sign of the effect changes at $\theta_{\rm L}=\theta_{\rm{M}}$.  This will be experimentally verified in the next section. The size of the shifts in Li are on the order of 100~kHz to 1~MHz, large enough to completely overshadow effects associated with Doppler shifts and optical pumping.

\begin{figure}[h]
\centering\includegraphics [width=3.3 in,angle=0] {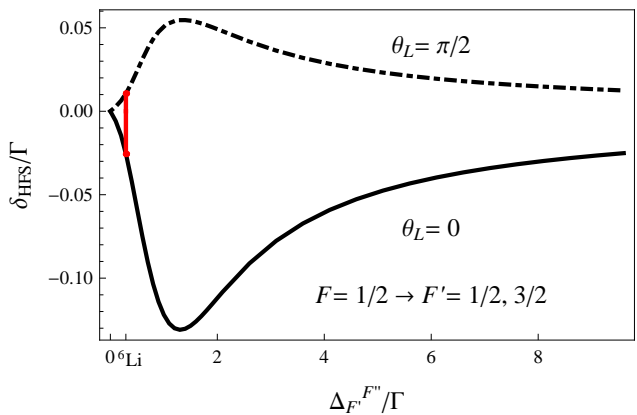}
\caption{(color online).
Error in the measured hyperfine splitting (in units of $\Gamma$) for a $F=1/2 \rightarrow F^\prime=1/2,3/2$ transition as a function of the assumed separation when the theoretically calculated full line shape is fit as the sum of two Lorentzians.  The dash dotted (solid) curve is for laser polarization $\theta_{\rm L}=\pi/2$~($0$).  The error in the hyperfine splitting is greatest where the assumed separation is about 1.3 times the natural line width.  The red vertical line indicates the $x$-position of the actual hyperfine splitting for $^6$Li.  Its vertical extent shows the range of errors that can occur when the laser polarization is varied between $0$ and $\pi/2$.}
\label{fig:ShiftsInHFCenters}
\end{figure}

To provide an estimate for other transitions not explicitly considered here, we imagine atoms with the electronic structure of $^{6}$Li or $^{7}$Li with variable hyperfine coupling. %$\vec{I}\cdot \vec{J}$,  enabling the hyperfine splitting to be varied.
We consider shifts of individual hyperfine components as the hyperfine splitting is varied. We intuitively expect that degenerate resonances would not affect the measured line position.  In the opposite limit, $\Delta^{F^{\prime \prime}}_{F^{\prime}}/\Gamma \gg 1$  we also expect the line positions to be unperturbed.  These two limits imply that there must be an intermediate hyperfine splitting that maximally affects the measured line positions. We can see in Fig.~\ref{fig:ShiftsInHFCenters} that this happens when $\Delta^{F^{\prime \prime}}_{F^{\prime}}/\Gamma$ is of order one.
%To get a feel for the scaling of individual component shifts as a function of separation, we consider a simple analytically solvable model of two equal amplitude Lorentzian distributions both with and without a cross-term. We compute the position of one of the two symmetric component centers by finding the zero of the first derivative of the distribution.  We then take the difference between a component center of the two Lorentzian distribution, $x_{\rm L}$, and a component center of the full line shape including cross-terms, $x_F$, as a function of resonance splitting $\Delta$.

To get a feel for the apparent shifts of individual components as a function of separation, we consider a simple analytically solvable line shape consisting of two Lorentzian profiles with splitting $\Delta$ and equal amplitude.  We take line profiles with and without cross terms and determine the component positions for each as the zero crossings of their first derivatives. We examine the difference of the position of the first component in the Lorentzian only profile, $x_{\rm L}$, and the position of the corresponding component in the full line profile including cross terms, $x_{\rm F}$, as a function of the splitting $\Delta$.  In the limit of distantly spaced resonances, $\Delta/\Gamma \gg 1$, the difference in line centers is $x_F - x_{\rm L} \simeq \Gamma^2/4 \Delta$, in agreement with the large splitting limit described in~\cite{Hessels2010}.  These shifts at large separation have recently been calculated at the 1~kHz level in meta-stable He~\cite{MarsmanAgainPRA2012} and in principle occur in muonic hydrogen, although at $\approx$100~MHz they are much too small to account for the discrepancy between proton charge radius values~\cite{PohlNature2010,MohrCODATA}.  In alkalis with resolvable hyperfine structure, i.e. $^{87}$Rb and $^{133}$Cs, these shifts may also appear at the $\approx$10~kHz level which, while much smaller than in unresolvable lines is on the order of the reported experimental uncertainties \cite{YeOpticsLett1996,GerginovCsD2PRA2004}.  This zero intensity shift may also arise from fine structure interference, and for Li is $\approx$860~Hz~(below our experimental uncertainty).  These shifts at large separation may be particularly insidious because they would only add a weak linear dependance to the background without deforming the line shape as in the case of unresolvable features.

%where $x_F$ is a component peak of the full line shape including the cross-terms and $x_{\rm L}$ is a component peak of the Lorentzian line shape excluding cross-terms,

\begin{figure}[h]
\centering\includegraphics [width=3.3 in,angle=0] {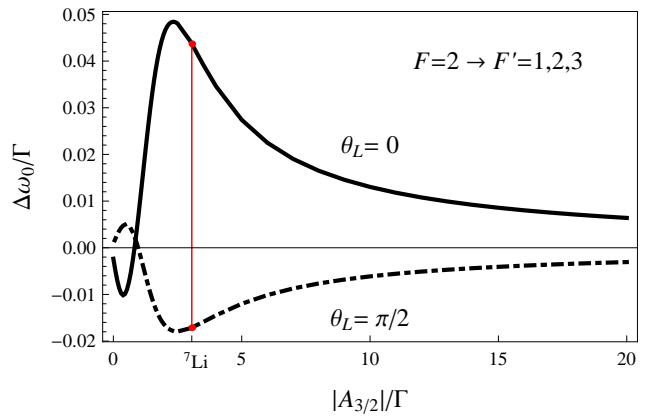}
\caption{(color online).
Error in the measured center of gravity (in units of $\Gamma$) for a  $F=2 \rightarrow F^\prime =1,2,3$ transition as a function of the hyperfine A constant when the theoretically calculated full line shape is fit as a sum of three Lorentzians. The dash dotted (solid) curve is for laser polarization $\theta_{\rm L}=\pi/2$~($0$).  The red vertical line indicates the $x$-position of the actual $A$ constant for $^7\rm{Li}$.  Its vertical extent shows the range of errors that can occur when the laser polarization is varied between $0$ and $\pi/2$.}
\label{fig:ShiftsInCentersofGravity}
\end{figure}

We also investigate the dependence of an unresolved feature's extracted center of gravity on hyperfine separation as shown in Fig.~\ref{fig:ShiftsInCentersofGravity}.  Using the same procedure, we generate the full line shape,  now with three components~(F=2 $\rightarrow$ F'=1,2,3).
%Since there are 3 components their relative spacing is controlled by adjusting
We vary the splitting via the magnetic dipole constant,~A$_{3/2}$,  while fixing the electric quadrupole constant at the value appropriate for $^7$Li.  The same qualitative behavior occurs, producing extracted center of gravity shifts which are largest when $|A_{3/2}| (\propto \Delta^{F^{\prime \prime}}_{F^{\prime}})$ is of order $\Gamma$.  There is now an additional feature, since there are two resonances that can shift relative to each other, the sign of the shift can change for a given laser polarization.

%To do: theoretical lines shape with second unequal amplitude ``b"

\section{Application to $^{6,7}$Li Experimental Data}
\label{sec:experiment}
\begin{figure}[h]
\centering\includegraphics [width=3.2 in,angle=0] {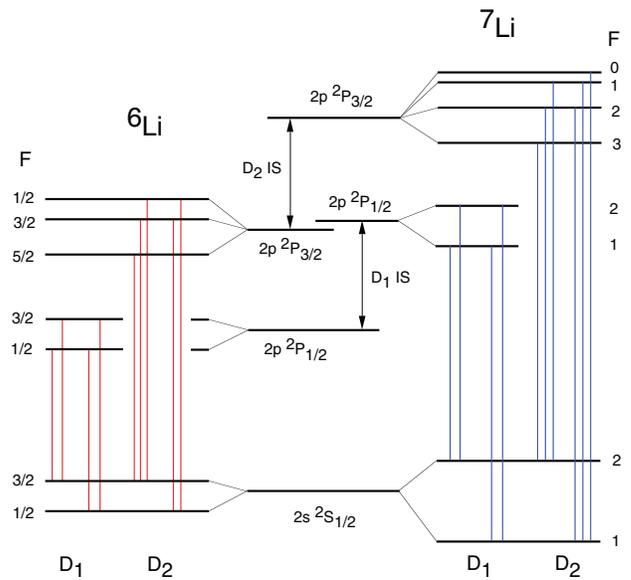}
\caption{(color online).  Relevant $^{6,7}$Li level structure. The hyperfine components for the D2 transition have natural widths of order the hyperfine splitting.  IS~(isotope shift)}
\label{fig:Grotrain}
\end{figure}

Having discussed the nature and theoretical implications of quantum interference effects on the observed line shape, we apply our theoretical results to experimentally measured spectra of lithium taken at multiple laser polarization angles.  Improved spectroscopy of the Li D lines, see Fig.~\ref{fig:Grotrain} for level structure, is of broad interest in physics because the isotope shift of these lines may serve as a nuclear-model-independent method to measure relative nuclear charge radii, which are especially interesting in the neutron rich $^{8,9,11}$Li~\cite{YanDrake2000}. Measured isotope shifts for the lithium 2s-2p~(D~lines)~\cite{Sansonetti1995,Scherf1996,Walls2003,Noble2006,Das2007} or 2s-3s~\cite{Lien2011,Sanchez2009,BushawPRL2003,Ewald2004} transitions can be combined with precise theoretical calculations~\cite{Nortershauser2011,Yan2008,YanDrake2000} to determine relative nuclear charge radii of lithium isotopes.  Additionally, measured D-line transition frequencies are used as input for the calculation of species-specific ``tune in/out" optical lattices for mixtures of quantum degenerate gases~\cite{LeBlancPRA2007,SafPRA2011,SafronovaPRA2012}.

Our additional measurement and analysis provides a refined determination of the
absolute transition frequencies of the $^{6,7}$Li D2 lines.  When combined with previously measured D1 values~\cite{Sansonetti2011, *Sansonetti2011erratum} these new data provide an improved measure of the $^{6,7}$Li excited state fine structure, 2s-2p isotope shift, and the isotopic difference in the $^2$P fine-structure splitting, the splitting isotope shift~(SIS).  The SIS provides the best point of comparison between theory and experiment.
%and the best point of comparison between theory and experiment, the isotopic difference in the fine structure splittings also called the splitting isotope shift~(SIS).
We propose that the interference effect we describe here is the root cause for some disagreements between previous measurements in Li~\cite{Sansonetti1995,Scherf1996,Walls2003,Noble2006} and for the lack of internal consistency of the frequency comb based measurement in K~\cite{Falke}.

%Having theoretically discussed the nature and implications of the full line shape, we apply it to experimentally obtained Doppler broadened spectra taken at multiple laser polarization angles to offer a refined measurement of the $^{6,7}$Li isotope shift. We believe that this effect is the root cause for disagreements between multiple measurements in Li\cite{Sansonetti1995,Scherf1996,Walls2003,Noble2006,Das2007} and for the lack of internal consistency frequency comb based measurement in K\cite{Falke}. We supplement our low intensity scattering theory with numerical simulations to account for the finite excitation laser power. We also improve our previously reported uncertainty in the magic angle offset using a combination of previously unavailable experimental input and further theoretical analysis.

%Improved spectroscopy of Li is of broader interest in physics because it may serve as a nuclear model independent method to measure nuclear charge radii which are especially interesting in the more neutron rich $^{8,9,11}$Li.  Measured isotope shifts for the lithium 2s-2p (D lines)\cite{Sansonetti1995,Scherf1996,Walls2003,Noble2006,Das2007} and 2s-3s\cite{Lien2011} transitions could be combined with precise theoretical calculations\cite{Nortershauser2011,Yan2008,YanDrake2000} to determine relative nuclear charge radii of lithium isotopes.

\subsection{Apparatus and procedure}
\label{apparatus}
\begin{figure}
\centering\includegraphics[width=3.0in,angle=0]{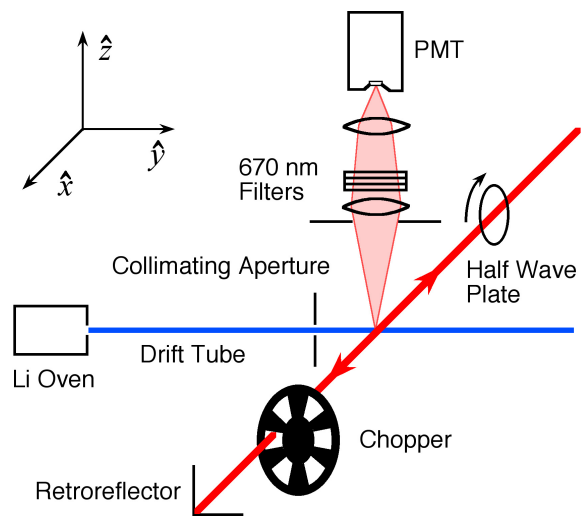}
\caption{\label{FigAppratus}Simplified schematic diagram of the experimental apparatus. The interaction region is surrounded by three layers of mu-metal (not shown) to minimize the magnetic field. The coordinate system shown is consistent with Fig.\ref{fig:coordinatesystem}}
\end{figure}

A simplified schematic view of our apparatus~\cite{Simien2011,Sansonetti2011, *Sansonetti2011erratum} is shown in Fig. \ref{FigAppratus}.  Light from a single frequency diode laser intersects a collimated thermal beam of lithium atoms at a right angle.  A half wave plate controls the angle of polarization of the light.  The laser beam is retroreflected by a precise corner cube that provides a return beam anti parallel to better than 1.45~$\mu$rad. The return beam is chopped at 500 Hz by a mechanical chopper.  We observe the spectrum by scanning the laser frequency over a lithium component and record the fluorescence along an axis approximately orthogonal to both the laser and atomic beams. To minimize stray light, the interaction region is imaged on the photocathode through a stack of three narrow band 670 nm interference filters.

The lithium beam is formed in a vacuum system with a background gas pressure of less than 1.3x10$^{-5}$ Pa (1x10$^{-7}$ Torr).  Lithium atoms effuse from an oven that is typically operated at 450 $^{\circ}$C and are collimated to a beam with a divergence angle of 1.4 mrad by a 2 mm aperture at a distance of 1.4 m.  Isotopically enriched $^6$Li was added to the oven to produce a beam with approximately equal densities of the two naturally occurring isotopes.

The lithium resonances are probed by a diode laser at 670 nm that is locked to an evacuated Fabry-Perot cavity using the Pound-Drever-Hall method \cite{Drever1983}.  This servo-lock narrows and stabilizes the diode laser output.  Despite the wide bandwidth of the servo, the laser line width is limited to about 500 kHz due to acoustic noise that couples to the cavity.  The laser can be scanned under computer control by varying the voltage applied to a piezo electric stack to which one of the cavity mirrors is mounted.  In the interaction region the laser is collimated to a 3.5 mm diameter beam and the laser power was typically attenuated to 3 ${\mu}$W.  Stability of the laser power over a single scan was better than 1\%.

The lithium fluorescence signal is detected in two channels by a gated photon counter.  One of these channels observes the fluorescence when both forward and return laser beams interact with the lithium beam.  For the other channel the return beam is blocked by the chopper and the signal is attributable to the forward beam only.  By differencing the photon count in the two channels, we recover the signal due to the reverse beam.  In this way we obtain the forward and reverse signals simultaneously in a single scan with an optical setup in which the anti-parallelism of the forward and reverse beams is limited only by the precision of the corner cube retroreflector.

Our experiment differs from all previous observations of the lithium D lines in that we measure directly the frequency of the laser using a femtosecond optical frequency comb\cite{Udem2002}.  The comb is a commercial instrument based on an Er fiber laser with a repetition rate of 250 MHz.  The fiber laser output is frequency doubled and broadened with a photonic crystal fiber producing a comb with broad spectral coverage in the red and near infrared regions.  A low resolution spectrometer is used to observe the spectral distribution of the comb to optimize the output at 670 nm.  The repetition rate and carrier offset frequency of the comb are referenced to a stable quartz oscillator which is in turn locked to a cesium clock.  This configuration produces a frequency reference with an absolute accuracy of better than 2 parts in 10$^{13}$ and an Allan deviation of approximately 3x10$^{-13}$ for integration times of 1 s to 100 s.  The frequency measurement using the comb is, therefore, a negligible contributor to our experimental uncertainty.

The spectroscopy laser is beat against a single tooth of the frequency comb using a high speed photodetector and a narrow band filter having a center frequency of 30 MHz and a width of about 6 MHz.  To record a calibrated scan across a lithium line, the repetition rate of the frequency comb is first adjusted so that the beat frequency between an arbitrary mode of the comb and the spectroscopy laser is approximately 30 MHz.  A computer generated voltage ramp is then used to vary both the laser frequency and the comb repetition rate so that the beat frequency remains fixed at 30 MHz.

Data are recorded by scanning the laser across a lithium resonance in steps of approximately 250 kHz.  A settling time of 200 ms is allowed after each step.  Scans are acquired in pairs with increasing and decreasing laser frequency.  Fluorescence data are accumulated alternately on the two gated photon counter channels for a total acquisition time of 72 ms on each channel.  The beat note frequency between the spectroscopy laser and the frequency comb is counted over the same time interval.  For every data point the comb repetition rate, comb offset frequency, beat note frequency, beat note signal strength, lithium fluorescence signal on both photon counter channels, and spectroscopy laser output power are recorded.

Doppler free spectra of the Li D lines were taken at different laser polarization angles $\theta_{\rm L}$ and fit using the line shapes presented here convolved with a Gaussian to account for the residual Doppler broadening present in the experiment, typically $\approx 4$~MHz.~%(fits give 2 MHz)
For resolved resonance features without a polarization dependence, such as the D1 lines, the independent fitting parameters are the line center, the overall amplitude, a constant background offset, the natural width, and the Doppler width.  The polarization angle of any given data set was fixed.
%The zero offset angle was parametrically varied over a few degrees to optimize self consistency and is further discussed in subsection ~\ref{systematics}.
For the unresolved fluorescence features, we limited the number of fitting parameters by fixing the excited state hyperfine splittings to values calculated in~\cite{Puchalski2009} and in agreement with~\cite{Yerokhin2008}. In addition we fixed the ratio of the unresolved amplitudes to values given by Eq.~\ref{eq:fandgoftheta}, with numerical values for $A^{F'}_F$, $B^{F'}_F$, and $C^{F',F''}_F$ tabulated in Appendix~\ref{sec:tablesappendix}.  A small correction was made to account for the effect of the finite collection angle of the detector~(see Appendix~\ref{sec:NAappendix}).

\subsection{Observation of apparent line-strength and transition frequency variation with $\theta_{\rm L}$}
\label{observedshifts}
\begin{figure}[h]
\centering\includegraphics [width=3.3 in,angle=0] {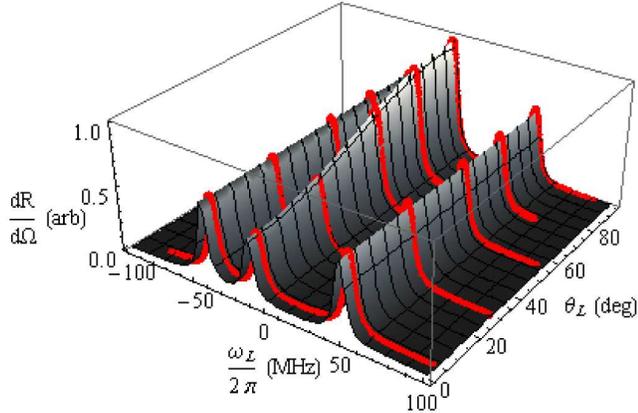}
\caption{(color online). Amplitude of scattered light, proportional to Eq.~\ref{eq:lineshape3}, as a function of laser frequency $\omega_{\rm L}$ and laser polarization angle $\theta_{\rm L}$.  The laser frequency is offset from the $^6$Li $\rm{F}=1/2$ ground state by 446~THz~(see table~\ref{Table1} for optical frequencies). The gray scale surface is the complete theoretical line shape including cross-terms and the red points are experimental data taken at $\theta_{\rm L}=0^\circ,25^\circ,51^\circ,75^\circ \rm{and}~ 90^\circ$.  The central feature is the $F=1/2 \rightarrow F^\prime=1/2,3/2$ transitions in $^6$Li while the two constant amplitude side peaks are the $\rm{F}=1 \rightarrow \rm{F'}=1,2$ D1 lines of $^7$Li.}
\label{fig:Li6D2HighSurfacePlot}
\end{figure}

One of the most striking features present in the more complete line shapes is the change in the amount of scattered light with excitation polarization.  A single fit to five spectra at different laser polarization angles demonstrates good overall agreement, including relative line-strengths. Fig.~\ref{fig:Li6D2HighSurfacePlot} shows the $F=1/2 \rightarrow  F^\prime =1/2,3/2$ D2 feature of $^6 {\rm Li}$~(center, $\omega_{\rm L}/2\pi\approx$0~MHz) and the $ F=2 \rightarrow F^\prime =1,2$ D1 peaks of $^7\rm{Li}$ (left and right, $\omega_{\rm L}/2\pi\approx\pm50$~MHz).  The $^7\rm{Li}$ D1 lines have no angular dependence (in general no D1 lines have angular dependence). The presence of the D1 lines enable the single fit to multiple data sets because they allow the effect of background light levels and laser intensity fluctuations to be compensated for in multiple spectra taken at different times.  The fit to these five data sets used only one natural width and one (mass scaled) Doppler width.

\begin{figure}[h]
\centering\includegraphics [width=3.3 in,angle=0] {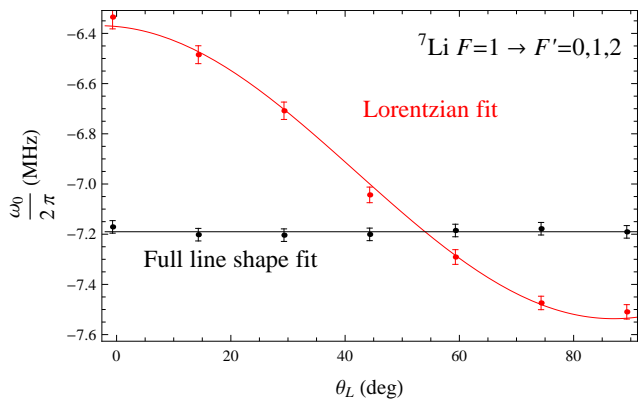}
\caption{(color online).   Line center of the $^7$Li $F=1 \rightarrow F^\prime=0,1,2$ feature fit from experimentally measured spectra as function of laser polarization angle with respect to the collection direction.  Transition frequencies are offset from the $^7$Li F=1 ground state by 446~THz~(see table~\ref{Table1} for optical frequencies). The black~(red) data points were extracted by fitting the data to functions with~(without) interference cross-terms. Error bars represent the uncertainties given in table~\ref{Table1}. }
\label{fig:CentersofAngle}
\end{figure}

To demonstrate the apparent transition frequency shifts resulting from analysis with an incomplete line shape in measured $^7\rm{Li}$ D2 data, we fit the same spectra taken at different laser polarizations and extract the line centers, with and without the cross-terms. In Fig.~\ref{fig:CentersofAngle}, the red points are line centers fit without cross-terms and the black points are the same data fit with the full theory. The black points are self consistent, independent of laser polarization while the red points exhibit a strong polarization dependence. The fit to the red data is of the form $A+B P_2(\cos{\theta_{\rm L}})$.
%because the size of the missing terms is proportional to $P_2(\cos{\theta_{\rm L}})$.
The amplitude of the laser polarization dependent shift is of order 1~MHz.
 % and is somewhat larger than the Doppler-free numerical analysis in Fig~\ref{fig:ShiftsInCentersofGravity}.
%As visible in the error bars, the quality of the Lorentzian only fit improves near the magic angle.
Near the magic angle $\theta_{\rm L}=\theta_{\rm{M}}\approx 54.7$ the Lorentzian fits give the same linecenter as the full line shape.

\subsection{Discussion of Systematics}
\label{systematics}

\textit{Angular offset: }To accurately extract line positions at all polarizations, the angle $\theta_{\rm L}$ between the laser polarization and the detection optics must be controlled and understood.
%While most parts of the apparatus were carefully engineered, (e.g. the atomic beam was precisely orthogonal to the laser beam to suppress Doppler shifts to as low as $\approx$1 kHz in resolved features), the laser polarization was not a priori optimized for absolute postion accuracy.
Using a waveplate, we could precisely define $\theta_{\rm L}$ up to a small unknown offset angle $\theta_0$. We improved our previous estimate of $\theta_0$~\cite{Sansonetti2011, *Sansonetti2011erratum} by geometric measurements made when disassembling the apparatus, finding $\theta_0=-0.7(10)$~degrees.

As a consistency check, we compared the well known ground state hyperfine intervals~(GHI) to GHI values we measure by subtracting optical frequencies at multiple angles $\theta_{\rm L}$.  We note that for small offsets $\theta_0$, the line shifts near $\theta_{\rm L}=0,\pi/2$ are insensitive to first order in $\theta_0$ because the derivative of the angular dependence~($ \propto \sin \theta_{\rm L} \cos \theta_{\rm L}|_{\theta_{\rm L}=\gamma,~\theta_{\rm{s}}=0}$) vanishes.  This is of practical utility since data fit at $\theta_{\rm L}=0,\pi/2$ with the complete line shape including cross-terms should be accurate as well as equal to each other.  We found that while the GHI's derived from measurements of the resolved D1 lines \cite{Sansonetti2011, *Sansonetti2011erratum} were consistent with known values~\cite{Beckmann1974}, the GHI's derived from the unresolved D2 lines at $\theta_{\rm L}=0,\pi/2$ differed from the known values by as much as 30~kHz.  This disagreement indicates the importance of intensity dependent shifts on the fitted line shapes when cross-terms are significant. %which we account for below.

%In addition, the $\theta_0$ that best determined the ``magic'' angle matching the known GHI's did not agree with our geometric measurements by $\approx 5$ degrees.

\textit{Intensity dependent shifts: } For isolated lines, the fitted amplitudes are taken to be free parameters and the fitted line centers are independent of fitted amplitude. As a consequence, the centers of the resolved lines are not sensitive to intensity dependent effects like optical pumping that modify the line ratios from their theoretical values. For unresolvable lines, however, the fitted line positions depend on the fixed relative values of $f$ and $g$ used in the fit.   The unresolvable lines are therefore sensitive to intensity dependent effects. To explore the impact of excitation laser intensity  on extracted line centers, we measured a subset of spectra at multiple laser powers and performed a full optical Bloch equation~(OBE) simulation of the scattering, including all the ground and excited Zeeman levels~\cite{CohenTannoudjiAPI,QOToolboxTanJPhysB1999}. We numerically solve the OBE with a time-dependent Rabi frequency proportional to the Gaussian intensity profile seen by the atom as it transverses the excitation laser beam.  We then generate a Doppler free line shape by calculating the directional photon scattering rates derived from the OBE, as a function of laser frequency.
%We then numerically generate a Doppler free line shape by calculating the population after a Gaussian pulse as a function of laser frequency.
%at fixed excitation laser frequency, and determine a scattered line shape, accounting for optical pumping and ac Stark shifts.
At the intensities used here and in~\cite{Sansonetti2011, *Sansonetti2011erratum}, we find these intensity dependent effects are small but important~($\approx 20$ kHz). However, we suggest that larger previously reported uncertainties~($\approx$~100~kHz) in $^{39,41}$K~\cite{Falke} ascribed to optical pumping could likely be removed by using a line shape that includes crossterms.

To quantitatively account for intensity dependent light shifts and optical pumping effects on the line positions, we generate numerical OBE data at several different intensities and fit the numerical data using the analytically calculated line strengths $f$ and $g$ appropriate for low intensity. (We confirm that in the low intensity limit, the numerical data matches both the expected line positions and line strengths.) We then determine the linear intensity-dependent line shifts from this numerical data, and apply this shift to the measured line positions \footnote{Another approach might be to fit the numerical data with free line weights and use the determined intensity dependent line weight ratios in the experimental fits. This provides a large number of free parameters, however, and the fits to numerical data were unstable in some cases.}. The laser intensities were determined experimentally from the relative line strengths of the resolved features taken at different laser intensities.  This estimate of the intensity is somewhat lower than estimates based on measured beam waists and laser power~(typically 3.5~mm and 3~$\mu$W respectively) but removes uncertainty associated with secondary measurements of beam waist and power.  For most features, the shift was of order a few kHz/$\mu$W, but for the $^7$Li D2 $F=1 \rightarrow F^\prime = 0, 1, 2$ transitions it was as large as 6.7~kHz/$\mu$W~(for our beam waist).  The uncertainty in this correction was set equal to the value of the applied shift and represents one of the largest sources of uncertainty in the experiment. For the unresolvable lines considered here, we find that optical pumping can have a larger systematic effect than the light shifts alone.  Future experiments should be careful to work at low intensities to avoid these shifts on unresolvable lines.

\textit{Doppler correction: }The correction of the first order Doppler effect was determined from simultaneously recorded forward and reverse beam signals using a corner cube to retro-reflect the excitation laser beam.
For the polarization independent D1 lines \cite{Sansonetti2011, *Sansonetti2011erratum} the systematic contribution to the uncertainty of this correction is 1.4~kHz due to imperfections of the corner cube retroreflector.
%The small uncertainty in the Doppler contribution for the resolved lines of about 1.4~kHz is the systematic uncertainty due to imperfections of the corner cube retroreflector.
Because the retroreflector does not preserve the laser polarization, the Doppler correction for the polarization sensitive unresolved D2 lines could not be determined using the corner cube, and is taken instead from a linear fit of correction versus time for resolved components measured on the same day. This is necessary because the laser alignment drifts slightly over hours of data taking, and results in a larger Doppler uncertainty of about 10~kHz.

\subsection{Results: Absolute transition frequencies, excited state hyperfine splitting, isotope shift, and splitting isotope shift}
\label{Results}
\begin{figure}[h]
\centering\includegraphics [width=3.3 in,angle=0] {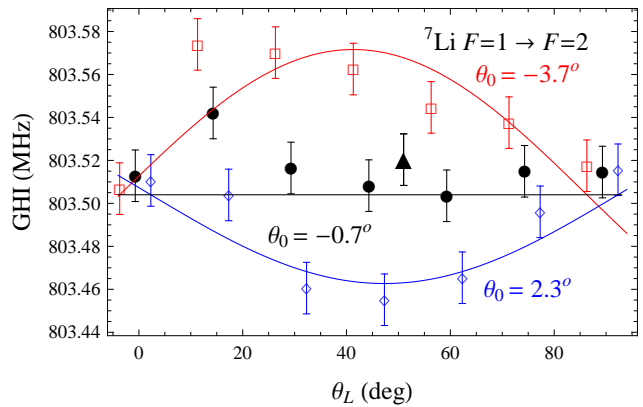}
\caption{(color online).  $^7$Li ground state hyperfine interval~($F=1 \rightarrow F=2$) as function of laser polarization angle. The measured GHI was determined by subtracting absolute measurements of the excited state $F=1 \rightarrow F^\prime=0,1,2$ and $F=2 \rightarrow F^\prime=1,2,3$ features including Doppler and intensity dependent corrections~(see the text).  The red~(black, blue) data is for an angular offset of $-3.7^\circ$~($-0.7^\circ$,$2.3^\circ$).  The red~(black,blue) curve is of the form $A_{\theta_0} \sin(\theta) \cos(\theta) + \rm{GHI_0}$, where $\rm{GHI_0}$ is the value measured in~\cite{Beckmann1974} and $A_{\theta_0}$ is fit to the data.
%The dashed line is the mean value of the data points at zero offset.
The triangular point is data from \cite{Sansonetti2011, *Sansonetti2011erratum} re-analyzed using the procedure described in the present work.  Error bars represent the uncertainties given in table ~\ref{Table1}.}
\label{fig:GSI}
\end{figure}

\begin{figure}[h]
\centering\includegraphics [width=3.3 in,angle=0] {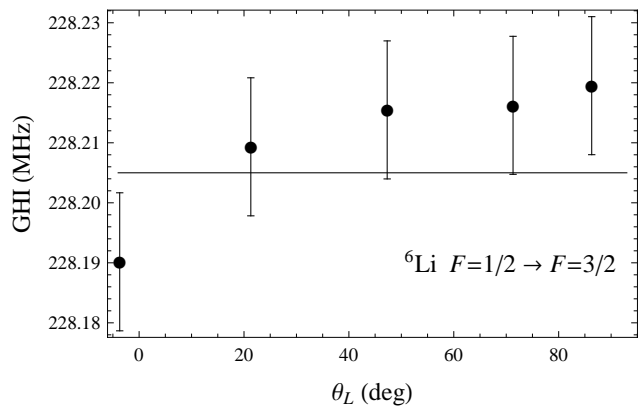}
\caption{(color online).  $^6$Li ground state hyperfine interval ($F=1/2 \rightarrow F=3/2$) as function of laser polarization angle. The measured GHI was determined by subtracting absolute measurements of the excited state $F=1/2 \rightarrow F^\prime=1/2,~3/2$ and $F=3/2 \rightarrow F'=1/2,~3/2,~5/2$ features. The solid line is the value measured in~\cite{Beckmann1974}.  Error bars represent the uncertainties given in table ~\ref{Table1}. %and dashed line is the mean value of the data points at zero offset
}
\label{fig:GSI6}
\end{figure}

Including the Doppler corrections and the power dependent shifts, the GHI values at $\theta_{\rm L}=0,\pi/2$ are in agreement with each other and the known values~\cite{Beckmann1974}~(see Figs.~\ref{fig:GSI}~and~\ref{fig:GSI6}).  The value of $\theta_0$ that minimizes the $ \sin \theta_{\rm L} \cos \theta_{\rm L}$ angular dependence is consistent with the geometrically determined value.  The final reported line positions, shown in table~\ref{Table1}, represent an average over $\theta_{\rm L}$.  A representative uncertainty budget is given in table~\ref{Table2}.

Measurements at multiple laser polarizations analyzed with the correct line shape provide an important tool to independently estimate systematic errors associated with the offset angle $\theta_0$. For example, power-dependent shifts such as optical pumping can partially cancel the effect of $\theta_0$ on the line shape and GHI. Minimizing residuals and comparing the GHI near $\theta_m$ can still lead to small systematic shifts in the line positions. These effects are more prominent in $^7$Li than $^6$Li, and our new determinations of the absolute cog transition frequencies differ from our previous results~\cite{Sansonetti2011, *Sansonetti2011erratum}, by 83~kHz and 19~kHz, respectively. From the absolute frequencies the excited state fine structure splitting~(Table~\ref{Table3}), as well as the 2s-2p IS and the SIS~(Table~\ref{Table4}) are calculated and compared to the existing literature.
As discussed in \cite{Yan2008}, both quantum electrodynamic and nuclear size corrections largely cancel when calculating the SIS.  It is, therefore, the most reliable result of theory and has been suggested as a benchmark for testing the internal consistency of experimental data.  Previously reported results have disagreed with each other and with theory far beyond their reported uncertainties (Table~\ref{Table4}).  Our current result resolves these discrepancies and is in full agreement with the most recent theoretical result~\cite{Puchalski2009}.  This supports the theory that underlies the use of D-line IS's to determine mean square nuclear charge radii for short lived Li isotopes.
 %The most sensitive comparison between theory is the SIS which is in good agreement with best reported theory~\cite{Puchalski2009}. This establishes the potential viability of D line measurements in short lived isotopes.

%We use the new values for the unresolvable D2 lines to compute the excited state fine structure splitting~(table~\ref{Table3}), as well as the 2s-2p IS and the SIS=$^7$Li~IS$- ~^6$Li IS~(table~\ref{Table4}).

%[Note: individual component values in the table have not been updated, just the cog's]
\begin{table}
\caption{ \label{Table1} Measured frequencies of hyperfine components and centers of gravity~(cog) of the  $^{6,7}$Li D2 lines.\footnote{Unresolved component splittings for the D2 lines are fixed at values calculated from the hyperfine A and B constants of~\cite{Puchalski2009}}}
\begin{ruledtabular}
\begin{tabular}{lccd}
Line & $F$ & $F^\prime$ & \multicolumn{1}{c}{\textrm{Frequency (MHz)}} \\
\colrule

$^6$Li D2 & 3/2 & 5/2 & 446799571.067(21)\\
          & 3/2 & 3/2 & 446799573.962(21)\\
          & 3/2 & 1/2 & 446799575.673(21)\\
          & 1/2 & 3/2 & 446799802.172(16)\\
          & 1/2 & 1/2 & 446799803.883(16)\\
\cline{4-4} \vspace{1.5mm} $^6$Li D2 cog & &     & 446799648.870(15)\\
$^7$Li D2 & 2 & 3 & 446809874.895(20)\\
          & 2 & 2 & 446809884.357(20)\\
          & 2 & 1 & 446809890.170(20)\\
          & 1 & 2 & 446810687.873(25)\\
          & 1 & 1 & 446810693.687(25)\\
          & 1 & 0 & 446810696.445(25)\\
\cline{4-4} $^7$Li D2 cog & & & 446810183.163(16)\\
\end{tabular}
\end{ruledtabular}
\end{table}

\begin{table}
\caption{\label{Table2}Representative uncertainty budget~(kHz).}
\begin{ruledtabular}
\begin{tabular}{ld}
Uncertainty  & \multicolumn{1}{c}{\textrm{$^6$Li D2}}\\
Component    & \multicolumn{1}{c}{\textrm{$F=3/2 \rightarrow F^\prime=5/2,3/2,1/2$}}\\
\colrule
Statistical variation  & 4 \\
First order Doppler effect  & 10\\
Estimate of ${\theta}_m$  & 3 \\
Laser power dependent shifts\footnote{Optical pumping, multiple excitation recoil, AC Stark shift}  & 17 \\
Laser intensity variation  & 3 \\
Hyperfine constant inaccuracy  & 2 \\
Imaging system imperfections  & 2 \\
Magnetic field shift  & <1 \\
\vspace{1.5mm}Reference frequency  & 0.089 \\
Total  & 21 \\
\end{tabular}
\end{ruledtabular}
\end{table}

\begin{table}
\caption{\label{Table3} Excited state fine-structure intervals.}
\begin{ruledtabular}
\begin{tabular}{ldl}
Interval\footnote{\label{footD1}All D1 values are taken on the same apparatus and reported in~\cite{Sansonetti2011, *Sansonetti2011erratum}} & \multicolumn{1}{c}{Splitting (MHz)} & Reference \\
\colrule
%$^7$Li 2s $^2$S hfs & 803.?? & this work \\
%        & 803.493(14) & Sansonetti ~\cite{Sansonetti2011, *Sansonetti2011erratum} \\
%\vspace{1.5mm} & 803.5040866(10) & Beckmann ~\cite{Beckmann1974} \\
%$^6$Li 2s $^2$S hfs & 228.?? & this work \\
%          & 228.215(17) & Sansonetti ~\cite{Sansonetti2011, *Sansonetti2011erratum}\\
%\vspace{1.5mm} & 228.205261(3) & Beckmann ~\cite{Beckmann1974} \\
$^6$Li 2p $^2$P fs & 10052.779(17)  & this work\\
          & 10052.799(22) & Sansonetti ~\cite{Sansonetti2011, *Sansonetti2011erratum}\\
          & 10052.76(22) & Brog ~\cite{Brog1967} \\
          & 10052.044(91) & Walls ~\cite{Walls2003} \\
          & 10052.964(50) & Noble ~\cite{Noble2006} \\
          & 10052.862(41) & Das ~\cite{Das2007} \\
\vspace{1.5mm}& 10050.932(8)\footnote{\label{footTeb}The uncertainties reported in \cite{Puchalski2009} represent only the numerical uncertainty and do not include any estimate of the size of corrections not included in the calculations.} & Puchalski(theory) ~\cite{Puchalski2009} \\
$^7$Li 2p $^2$P fs & 10053.310(17) & this work \\
          & 10053.393(21) & Sansonetti ~\cite{Sansonetti2011, *Sansonetti2011erratum} \\
          & 10053.184(58) & Orth ~\cite{Orth1975} \\
          & 10052.37(11) & Walls ~\cite{Walls2003} \\
          & 10053.119(58) & Noble ~\cite{Noble2006} \\
          & 10051.999(41) & Das ~\cite{Das2007} \\
          & 10051.477(8) \ref{footTeb} & Puchalski(theory) ~\cite{Puchalski2009} \\
\end{tabular}
\end{ruledtabular}
\end{table}

\begin{table}
\caption{\label{Table4}$^{7,6}$Li isotope shifts.}
\begin{ruledtabular}
\begin{tabular}{lll}
Transition &  \multicolumn{1}{c}{Shift (MHz)} & Reference \\
\colrule
D2 IS & 10534.293(22) & this work \\
    & 10534.357(29) & Sansonetti ~\cite{Sansonetti2011, *Sansonetti2011erratum} \\
    & 10533.59(14) & Walls ~\cite{Walls2003} \\
    & 10534.194(104) & Noble ~\cite{Noble2006} \\
\vspace{1.5mm} & 10533.352(68) & Das ~\cite{Das2007} \\
SIS\footnote{\label{footD1}All D1 values are taken on the same apparatus and reported in~\cite{Sansonetti2011, *Sansonetti2011erratum}}& 0.531(24) & this work \\
    & 0.594(30) & Sansonetti ~\cite{Sansonetti2011, *Sansonetti2011erratum} \\
    & -0.67(14) & Walls ~\cite{Walls2003} \\
    & 0.155(60) & Noble ~\cite{Noble2006} \\
    & -0.863(79) & Das ~\cite{Das2007} \\
    & 0.396(9) & Yan(theory) ~\cite{Yan2008} \\
    & 0.5447(1) & Puchalski(theory) ~\cite{Puchalski2009} \\
\end{tabular}
\end{ruledtabular}
\end{table}

\section{Extraction of relative nuclear charge radii}
\label{sec:ncr}
\begin{figure}[h]
\centering\includegraphics [width=3.4 in,angle=0] {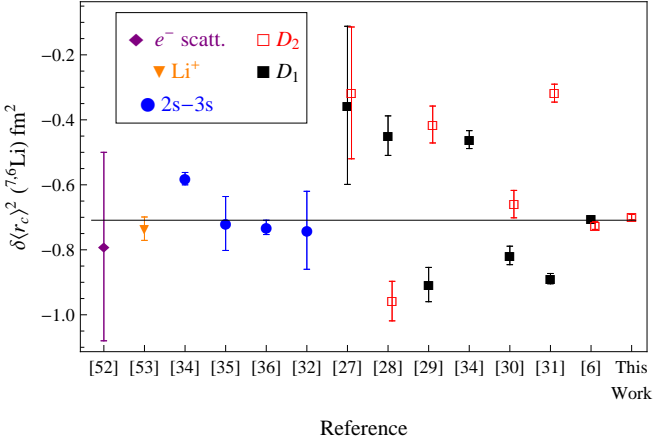}
\caption{(color online).
Measurements of the difference in mean square charge radius between $^{7}$Li and $^{6}$Li.  The points are grouped by type of measurement and are then ordered chronologically within different types of measurement.
The solid black line is the weighted average of the results of references~\cite{ElectronScatteringNCRDeJager1974,RiisPRA1994,Nortershauser2011}, along with the D1 value from~\cite{Sansonetti2011, *Sansonetti2011erratum} and the D2 value from this work.  Error bars for the present work represent the uncertainties given in table~\ref{Table4}, all other error bars represent the uncertainties given in the original references.}
\label{fig:NCR}
\end{figure}

Finally, we calculate the difference in the $^{6,7}$Li nuclear charge radii using the measured D2 isotope shifts reported in Table~\ref{Table4} and the D1 shifts reported in~\cite{Sansonetti2011, *Sansonetti2011erratum}.  This serves as a point of comparison amongst different types of measurements including elastic electron scattering~\cite{ElectronScatteringNCRDeJager1974}, optical isotope shift measurements on the $^3S_1\rightarrow^3P_{0,1,2}$ transition in Li$^+$~\cite{RiisPRA1994}, and optical isotope shift measurements of the 2s-3s, D1, and D2 transitions in neutral Li\cite{Sansonetti1995,Scherf1996,Walls2003,Noble2006,Das2007,Lien2011,Sanchez2009,Ewald2004,BushawPRL2003} as shown in Fig.~\ref{fig:NCR}.  We calculate the difference in nuclear charge radius using Eq.~(40) of~\cite{YanDrake2000},
\begin{equation}
\delta \langle r_{\rm c}^2 \rangle (^{7,6}{\rm Li})=\langle r_{\rm c}^2 \rangle (^7{\rm Li})-\langle r_{\rm c}^2 \rangle (^6{\rm Li})=\frac{(E_{\rm{meas}}-E_{\rm{0}})}{C_{\rm{0}}}
\end{equation}
where $\langle r^2_{\rm c} \rangle (^i{\rm Li})$ is the mean square nuclear charge radius of the $i^\mathrm{th}$ isotope in fm$^2$, E$_{\rm{meas}}$ is the measured isotope shift in MHz,  $E_{\rm{0}}=-10532.5682(-10532.0237)$~MHz is the theoretically calculated isotope shift excluding the finite size corrections for the D2(D1) transitions~\cite{PachuckiPC} and $C_{\rm{0}} = -2.4658$~MHz/fm$^2$~\cite{PachuckiPC}.
The values of the difference in mean square nuclear charge radius are $-0.705(3)$~fm$^2$ for the D1 and $-0.700(9)$~fm$^2$ for the D2 lines.  These values are self consistent and have the smallest uncertainties yet reported.  They bring the D-line measurements into full agreement with the best values from electron scattering and optical IS measurements on 2s-3s and $^3S_1\rightarrow^3P_{0,1,2}$ transitions in Li and  Li$^+$ respectively.
%consistent with $-0.731(22)$~fm$^2$ reported in \cite{Nortershauser2011} although they have reduced error bars.

\section{Conclusion}
%[This conclusion is more of a summary...  we should state some points: interference causes shifts, it also makes the lines more sensitive to optical pumping than resolved lines.]
We have reviewed low intensity scattering theory as it applies to the spectroscopy of alkali atoms with unresolvable hyperfine structures.  We find that the effects of light polarization and quantum interference alter the relative line strengths and quantitatively affect the extraction of transition frequencies from data, even in the low intensity limit.  Optical pumping effects at finite excitation power can further complicate the line shape, which we account for numerically.  This leads to a revised determination of the $^{6,7}$Li D2 line frequencies and splitting isotope shift.  We identify several species: H~\cite{Eikema2001}, $^{22,23}$Na~\cite{Gangrsky1998},  $^{39,40,41}$K~\cite{Falke}, and $^{221}$Fr~\cite{CocFr1985}, $^{7,9,11}$BeII~\cite{BeNucChargeRadiJPG2010}, and $^{25}$MgII~\cite{MukherjeeEPJD2005}, for which these complete line shapes will enable the next generation of measurements.

%It is interesting to note that the unresolvable transitions which most strongly exhibit this effect are for the reason the most difficult to sub-Doppler laser cool.  This limitation was recently overcome by moving to shorter wavelength ns $\rightarrow$ (n+1)p cooling transitions~\cite{McKayPhysRevA2011,DuartePhysRevA2011}. However, despite their narrower line width which facilitates lower Doppler temperatures these transitions also may exhibit this effect.
%This concept may also be extended to fermionic Alkali-earth and Alkali-earth like atoms with unresolved hyperfine structure in the $^1\rm{P}_1$ states such as $^{25}\rm{Mg}$, $^{87}\rm{Sr}$ and $^{173}\rm{Yb}$.

\begin{acknowledgments}
We thank W. D. Phillips and Peter J. Mohr for helpful discussions.
\end{acknowledgments}

\appendix
\section{}
\label{sec:CGappendix}
{\em Expressions for the normalized dipole matrix elements:}
The vector components of ${\bf A}$ are easiest to describe in the spherical vector basis $ A_q$ appropriate for $\sigma^+$, $\pi$ and $\sigma^-$ light, where
\begin{eqnarray}
A_1 & = &-(A_x+ i A_y)/\sqrt{2}\nonumber \\
A_0 &=&A_z\nonumber \\
A_{-1} &=&(A_x- i A_y)/\sqrt{2}.
\end{eqnarray}
Using the Wigner-Eckart theorem, the dipole matrix elements are given in terms of reduced matrix elements as
\begin{equation}
(D_{F m}^{F^\prime m^\prime})_q = \frac{\langle F^\prime  || {\bf D} || F \rangle }{\sqrt{2 F^\prime +1}} \langle F m; 1 q | F^\prime m^\prime \rangle,
\end{equation}
where $ \langle F m; 1 q | F^\prime m^\prime \rangle $ is the Clebsch-Gordan coefficient for adding $|F, m\rangle$ to $|1,q\rangle$ to get $|F^\prime, m^\prime \rangle$.  Under the assumption that the hyperfine interaction does not modify the electronic structure of the state, the $F$-reduced matrix elements can be written in terms of $J$-reduced elements
\begin{equation}
\langle F^\prime  || {\bf D} || F  \rangle = \langle J^\prime  || {\bf D} || J \rangle \sqrt{f_F^{F^\prime}},
\end{equation}
where the reduced oscillator strength $f^{F^\prime}_F$ for the $F$-$F^\prime$ transition can be written in terms of Wigner 6-j symbols:
\begin{equation}
\sqrt{f^{F^\prime}_F}= (-1)^{F+I+1+J^\prime} \sqrt{2 F +1} \sqrt{2 F^\prime +1} \left\{
\begin{array}{ccc}
    J^\prime    & J & 1   \\
    F    & F^\prime & I
    \end{array}
\right\}.
\end{equation}
Defining the components of the matrix elements $ ({\bf A}^{ F^\prime m^\prime}_{F m})_q$ for each $J\rightarrow J^\prime$ transition
\begin{equation}
({\bf A}^{ F^\prime m^\prime}_{F m})_q = \frac{\sqrt{2 J^\prime +1}}{\sqrt{2 F^\prime +1}} \langle F m; 1 q | F^\prime m^\prime \rangle \sqrt{f^{F^\prime}_F},
\label{eq:A5}
\end{equation}
the dipole matrix elements can be written as
\begin{equation}
(D_{F m}^{F^\prime m^\prime})_q = \frac{\langle J^\prime  || {\bf D} || J \rangle }{\sqrt{2 J^\prime +1}}  ({\bf A}^{ F^\prime m^\prime}_{F m})_q~.
\end{equation}
Pulling the reduced matrix element $\langle J^\prime  || {\bf D} || J \rangle$ out of the sum, Eq.~\ref{eq:HK} can be written in terms of the inverse scattering rate $\Gamma$ and a saturation intensity $I_0$,
\begin{equation}
\Gamma= \frac{1}{\tau} = \frac{\omega^3}{3 \pi \epsilon_0\hbar c^3}
\frac{\left| \langle J^\prime || {\bf D} || J \rangle\right|^2}{(2 J^\prime +1)},
\end{equation}
and
\begin{equation}
I_0 =\frac{  \pi h c  \Gamma} {3 \lambda^3},
\end{equation}
giving Eq.~\ref{eq:lineshape1}, where  $\omega$ and $\lambda$ are the frequency and wavelength of the transition.

\section{}
\label{sec:tablesappendix}
{\em Calculation of weights $f$ and $g$: }The dipole radiation weights $f$ and $g$ are calculated using the expression for ${\bf A}^{ F^\prime m^\prime}_{F m}$~(Eq.~\ref{eq:A5}) to determine $C^{F^\prime}_{i\rightarrow f}$~(Eq.~\ref{eq:4}), evaluating the sums in Eq.~\ref{eq:fandg} %~\ref{eq:7},
and then comparing to the dipole radiation pattern Eq.~\ref{eq:fandgoftheta}.  Taking $k_s$ along $\hat{z}$, ~(i.e. $\theta_{\rm{s}}=0$), with the two scattered polarizations $\hat{\epsilon}_{s1}=\hat{x}$ and $\hat{\epsilon}_{s2}=\hat{y}$,  and $\hat{\epsilon}_{\rm L}$ to lie in the $\hat{z}$-$\hat{y}$ plane as in Fig~\ref{fig:coordinatesystem}, the terms in the sum are given by
\begin{eqnarray}
\hat{\epsilon}_{s1}\cdot {\bf A}^{ F^\prime m^\prime}_{F m} & = &
 \frac{-1}{\sqrt{2}}\left(({\bf A}^{ F^\prime m^\prime}_{F m})_{1}
                -({\bf A}^{ F^\prime m^\prime}_{F m})_{-1}\right)\\
\hat{\epsilon}_{s2}\cdot {\bf A}^{ F^\prime m^\prime}_{F m}  & = &
 \frac{i}{\sqrt{2}}\left(({\bf A}^{ F^\prime m^\prime}_{F m})_{1}
              +({\bf A}^{ F^\prime m^\prime}_{F m})_{-1}\right) \\
\hat{\epsilon}_{L} \cdot {\bf A}^{ F^\prime m^\prime}_{F m} & = &
  \frac{i \sin{\theta_{\rm L}}}{\sqrt{2}}\left(({\bf A}^{ F^\prime m^\prime}_{F m})_{1}
            +({\bf A}^{ F^\prime m^\prime}_{F m})_{-1}\right) \nonumber\\
           & & + \cos{\theta_{\rm L}} \left({\bf A}^{ F^\prime m^\prime}_{F m}\right)_{0}.
\end{eqnarray}

We report line weights and cross-terms for the D2 transitions, $^2\mathrm{S}_{1/2} \rightarrow~ ^2\mathrm{P}_{3/2}$, of alkali atoms and hydrogen with $\rm{I}=1/2,1,$ and $3/2$ in tables\ref{Tabhalf},\ref{Tabone}, and \ref{Tabthreehalf} respectively.
%\usepackage{placeins}
%\FloatBarrier
%\begin{figure}[H]
\begin{table}[!ht]
\caption{\label{Tabhalf} D2 weights and cross-terms for $\rm{I}=1/2$ applicable to H, $^{11}$BeII }
\begin{ruledtabular}
\begin{tabular}{cccccc}
F & F' & $F"$ & $A^{F'}_F$  & $B^{F'}_F$ & $C^{F',F"}_F$ \\
\colrule
0 & 1 &   & 1/6  & -1/12  &   \\
1 & 1 & 2 & 1/12  & 1/48   & -1/16 \\
1 & 2 &   & 5/12  & -7/48   &   \\
\end{tabular}
\end{ruledtabular}
\end{table}

\begin{table}
\caption{\label{Tabone} D2 weights and cross-terms for $\rm{I}=1$  applicable to $^2$H and $^6$Li,$^{28}$Na }
\begin{ruledtabular}
\begin{tabular}{cccccc}
F & F' & $F"$ & $A^{F'}_F$  & $B^{F'}_F$ & $C^{F',F"}_F$ \\
\colrule
1/2 & 1/2 & 3/2 & 8/81  & 0        & -4/81  \\
1/2 & 3/2 &     & 10/81 & -1/81    &   \\
3/2 & 1/2 & 3/2 & 1/81  & 0        & 2/405   \\
3/2 & 3/2 & 5/2 & 8/81  & 16/2025  & -14/225   \\
3/2 & 5/2 & 1/2 & 1/3   & -7/75    & -1/90  \\
\end{tabular}
\end{ruledtabular}
\end{table}

\begin{table}
\caption{\label{Tabthreehalf} D2 weights and cross-terms for $\rm{I}=3/2$  applicable to $^{7,9,11}$Li,$^{21,23,34}$Na, $^{39,41}$K and $^{87}$Rb, $^{7,9}$BeII}
\begin{ruledtabular}
\begin{tabular}{cccccc}
F & F' & $F"$ & $A^{F'}_F$  & $B^{F'}_F$ & $C^{F',F"}_F$ \\
\colrule
1 & 0 & 1 & 1/24  & 0       &  0 \\
1 & 1 & 2 & 5/48  & -1/48   &  -1/32  \\
1 & 2 & 0 & 5/48  & 0       &  -1/48  \\
2 & 1 & 2 & 1/48  & 1/1200  &  1/160  \\
2 & 2 & 3 & 5/48  & 0       &  -7/120  \\
2 & 3 & 1 & 7/24 & -7/100  &  -7/400  \\
\end{tabular}
\end{ruledtabular}
\end{table}
Note that there is no angular dependence to the D1 terms, and therefore no dipole dependence~($B^{F'}_F = C^{F',F''}_F =0$ for D1). Also note that $C^{F',F''}_F = C^{F'',F'}_F$, physically this is because scattering through $F''$ is indistinguishable from scattering through $F'$ when the $F''$ and $F'$ are overlapped within the natural width.
%$\rm{I}=2$  applicable to $^{8}$Li $^{20,30}$Na\\
%$\rm{I}=5/2$  applicable to $^{221}$Fr\\
%$\rm{I}=3$  applicable to $^{22,26}$Na\\
%$\rm{I}=4$  applicable to $^{24}$Na $^{40}$K\\
\section{}
\label{sec:NAappendix}
{\em Collection optics correction: }If fluorescence is collected over all solid angle there is no polarization dependent modification to the line shape.  The equations given in the text are valid for light scattered into an infinitesimal solid angle.  Here we find the modification to the angular dependent part of the line weights and cross-terms due to the finite numerical aperture of the fluorescence collection optics.  For a given laser polarization $\hat{\epsilon}_{\rm L} $, we may integrate over the final scattering directions $\hat{k}_{\rm{s}}$ allowed by the collection optics~(parameterized by $\theta_{\rm{s}},\phi_{\rm{s}}$).
\begin{eqnarray}
\hat{\epsilon}_{\rm L} & = & \sin{\theta_{\rm L}}\ \hat{y}+ \cos{\theta_{\rm L}}\ \hat{z} \nonumber \\
\hat{k}_{\rm{s}} & = & \sin{\theta_{\rm{s}}} \cos{\phi_{\rm{s}}}\ \hat{x}+ \sin{\theta_{\rm{s}}} \sin{\phi_{\rm{s}}}\ \hat{y} + \cos{\theta_{\rm{s}}}\ \hat{z} \nonumber \\
\hat{\epsilon}_{\rm L}  \cdot \hat{k}_{\rm{s}} & = & \cos{\gamma}
\end{eqnarray}
Performing the angular integrations over the isotropic part, where $d\Omega_s=d\phi d\cos(\theta_{\rm{s}})$, we find
\begin{equation}
\iint _{0,0}^{2 \pi,\theta_{\rm{C}}} d\Omega_s = 2 \pi (1- \cos{\theta_{\rm{C}}}) \equiv S_0.
\end{equation}
The angle dependent dipole part is scaled by
\begin{eqnarray}
\iint _{0,0}^{2 \pi,\theta_{\rm{C}}} d\Omega_sP_2(\cos{\gamma}) &=&  \pi \cos{\theta_{\rm{C}}} \sin^2{\theta_{\rm{C}}} P_2(\cos{\theta_{\rm L}})   \nonumber \\
 & \equiv & S_2P_2(\cos{\theta_{\rm L}}).
\end{eqnarray}
Here $\theta_{\rm{C}}$ is the half angle of the fluorescence collection cone.  For determining experimentally relevant fitting functions, the ratio of the constant and dipole part is important, and we find that the dipole components are reduced relative to the constant components as,
\begin{equation}
S_2/S_0=\cos{\theta_{\rm{C}}} \cos^2{\left(\frac{\theta_{\rm{C}}}{2}\right)}.
\end{equation}
These scaling factors are included as part of the fitting functions to account for the numerical aperture of the imaging system.  Failure to include these scaling factors shifts the extracted line centers by $\approx 6$~kHz for $\theta_{\rm{C}}=26.6^{\rm{o}}$ used in this experiment.

%\section{}
%\label{sec:NucChargeRadius}

%\section{tables of applied corrections for internal use}
%\label{sec:tos}
%All shifts are in MHz.
%\\
%Doppler Corrections for Li 6 (From Aug 5 email) see table \ref{corr6}
%\\
%Power dependent shifts for 6 Li (From Feb 11 email)
%D2 high component
%+3.0 kHz/uW\\
%low component
%-6.4 kHz/uW
%\\
%Power dependent shifts for 7 Li see table\ref{corr7}\\
%D2 high component
%+6.4 kHz/uW\\
%low component
%-1.2 kHz/uW
%
%\begin{table}
%\caption{\label{corr7} $^{7}$Li corrections.}
%\begin{ruledtabular}
%\begin{tabular}{llllllll}
%CorrComp &   0 & 15 & 30 & 45 & 60 & 75 & 90 \\
%\colrule
%Doppler High & 0.131 & 0.135 & 0.138 & 0.141 & 0.145 & 0.151 & 0.155 \\
%Doppler Low & 0.186 & 0.182 & 0.178 & 0.174 & 0.170 & 0.165 & 0.162 \\
%Intensity High & 0.0192 & 0.0192 & 0.0192 & 0.0192 & 0.0192 & 0.0192 & 0.0192 \\
%Intensity Low & -0.0036 & -0.0036 & -0.0036 & -0.003 & -0.003 & -0.003 & -0.00264 \\
%\end{tabular}
%\end{ruledtabular}
%\end{table}
%
%
%
%\begin{table}
%\caption{\label{corr6} $^{6}$Li corrections.}
%\begin{ruledtabular}
%\begin{tabular}{llllll}
%CorrComp &   0 & 25 & 51 & 75 & 90 \\
%\colrule
%Doppler High & 0.258& 0.258& 0.260& 0.259& 0.260  \\
%Doppler Low & 0.262& 0.262& 0.261& 0.262& 0.263 \\
%Intensity High & 0.009& 0.009& 0.009& 0.009& 0.009 \\
%Intensity Low & -0.0192& -0.0192& -0.0192& -0.0192& -0.0192 \\
%\end{tabular}
%\end{ruledtabular}
%\end{table}

\bibliography{Libib}

\end{document}